\documentclass[english]{article}
\usepackage[T1]{fontenc}
\usepackage[latin9]{inputenc}
\usepackage[margin=1in]{geometry}
\usepackage{color}
\usepackage{array}
\usepackage{textcomp}
\usepackage{amsmath,amstext,amssymb}
\usepackage{graphicx,tikz,pgf}

\usetikzlibrary{arrows, automata, decorations.markings}
\tikzstyle{arrow}=[draw]

\tikzstyle{vecArrow} = [thick, decoration={markings,mark=at position
   1 with {\arrow[thick]{open triangle 90}}},
   double distance=3.0pt, shorten >= 4.5pt,
   preaction = {decorate},
   postaction = {draw,line width=1.4pt, white,shorten >= 1.5pt}]
\makeatletter

\providecommand{\tabularnewline}{\\}

\usepackage{tikz}\usepackage{url}
\usepackage{multicol}
\usepackage[english]{babel}

\makeatother

\usepackage{babel}
\begin{document}

\title{Circuit optimization for IBM processors: A way to get higher fidelity and higher values of nonclassicality witnesses}

\author{Mitali Sisodia$^{1}$,  Abhishek Shukla$^{2}$, Alexandre A. A. de Almeida$^{3}$, \\
 Gerhard W. Dueck$^{4}$, Anirban Pathak$^{1}$}

\maketitle
\begin{center}
$^{1}$Jaypee Institute of Information Technology, A 10, Sector 62, Noida,
UP 201307
\end{center}

\begin{center}
$^{2}$University of Science and Technology of China, Hefei 230026, P. R. China
\end{center}

\begin{center}
$^{3}$Department of Electrical Engineering, FEIS - Univ Estadual Paulista, Ilha Solteira, Brazil
\end{center}

\begin{center}
$^{4}$Faculty of Computer Science, University of New Brunswick, Canada
\end{center}

%
%
\begin{abstract}
Recently, various quantum computing and 
communication tasks have been implemented using IBM's superconductivity-based quantum
computers which are available on the cloud. Here, we show that the
circuits used in most of those works were not optimized and the use
of the optimized circuits can considerably improve the possibility of observing unique features of quantum mechanics.
Specifically, a systematic procedure 
is used here to obtain optimized circuits  (circuits having reduced 
gate count and number of levels) for a large number of Clifford+T circuits which
have already been implemented in the IBM quantum computers. Optimized circuits implementable in IBM quantum computers are also obtained for a set of reversible benchmark circuits. With a clear example, it is shown that the reduction in circuit costs enhances
the fidelity of the output state (with respect to the theoretically expected state in the absence of noise) as lesser number of gates and levels introduce lesser amount of errors during evolution of the state. Further, considering Mermin inequality as
an example, it's shown that the violation of classical limit is enhanced
when we use an optimized circuit. Thus, the approach adopted here can be used to identify relatively weaker signature of quantumness and also to establish quantum supremacy in a stronger manner.
\end{abstract}

\section{Introduction}

Since its introduction quantum computing has drawn considerable attention of the scientific community
because of the fact that it can perform certain computational tasks
much faster than its classical counter parts~\cite{grover1997quantum,shor1999polynomial}. For example,
it can search an unsorted database~\cite{grover1997quantum}, solve the discrete logarithm and prime factorization problems in a speed not achievable by  its classical counterparts~\cite{shor1999polynomial}. Similarly, quantum communication
has also drawn considerable attention, as it can perform classically
impossible tasks like teleportation~\cite{bennett1993teleporting} and as it can provide unconditional
security---an extremely desired feature that's not achievable in the
classical world (see~\cite{shenoy2017quantum} and references therein). These facts led to several theoretical proposals
for quantum computation and communication. Some of them have also
been verified experimentally. However, the access to experimental
facilities were restricted until the recent past and it was not available to most of the researchers. The
scenario has been changed considerably with the introduction of a
set of quantum computers by the IBM corporation~\cite{IBMQ}.
These quantum computers are placed in the cloud and researchers can access
them for free. Naturally, quantum computing research received a boost
with it, and several computational tasks (e.g., Bell state discrimination
\cite{sisodia2017experimental}, teleportation using optimal quantum resources
\cite{sisodia2017design}, quantum permutation algorithm~\cite{YZ2017optimization},
creation of a quantum check~\cite{BBP2017experimental}, testing Mermin inequalities
\cite{AL2016experimental}) have recently been realized using
IBM quantum computers. In these works, circuits formed using Clifford+T
gates have been used as IBM quantum computers allow only these gates.
However, no serious effort has yet been made to optimize the circuits. Although the
need for optimization was clear from the observations of various works~\cite{sisodia2017experimental}, where it was clearly
observed that the increase in gate count reduces the fidelity of the output state. In fact, there 
exists a direct (although not linear) correlation
between the gate count and the fidelity of the experimentally
obtained output state and the output state that would have
been  in the absence of noise.
Further, quantum process tomography of the gates used 
in IBM quantum computers has been performed by some of the present authors ~\cite{shukla2018complete}, and the same has  revealed that the 
gate fidelity of the gates used in IBM quantum computers are usually lower than the same obtained in
other technologies, like NMR. Because of these facts, any particular architecture of IBM quantum computers
imposes an upper-bound on the number of gates that can be used to construct an experimentally realizable 
quantum circuit with a reliable output. These limitations of the IBM quantum computers lead to the requirement that to obtain 
the best possible results using the IBM quantum computers, the number of gates
in circuits should be minimized. Keeping this requirement in mind, in a recent
work, some of the present authors designed an algorithm that can provide
optimized Clifford+T circuits~\cite{dueck_DSD_2018}. The same algorithm is used here to
obtain optimized circuits corresponding to a set of interesting Clifford+T
circuits that have been implemented using IBM quantum computers. In addition, a large number of reversible circuits are converted to Clifford+T circuits and optimized to show the usefulness of the approach used here and to establish that those circuits can also be implemented reliably using IBM quantum computers. Further, in what follows in the context of 
Mermin inequalities, it is shown that the optimized circuits not only provide better fidelity
it also enhances the amount by which the classical limit is violated. In other words, an optimized quantum circuit 
can more clearly reveal the nonclassical features, and thus help to strongly establish quantum supremacy. The beauty of the approach adopted here is that the method used is very general in nature and the optimization algorithm can be easily transformed from one architecture to another architecture of quantum computer.

The rest of the paper is structured as follows. Section~\ref{sec:How-to-optimize}
briefly describes the method adopted here for obtaining the optimized
quantum circuits. In Section~\ref{sec:Benefit-of-the-optimization},
we establish the benefits (advantages) of optimization through an  illustrative example. In Section~\ref{sec:More-results:-Optimized},
the work of previous section is extended to provide optimized circuits
for various quantum computing tasks. However, the improvement is quantified
only through the number of levels and gate count. Further, a large set of reversible circuits are also optimized. Finally, the paper
is concluded in Section \ref{sec:Conclusion}. 

\section{Optimized Clifford+T quantum circuits for IBM's QX2
and QX4 architectures\label{sec:How-to-optimize} }

\begin{figure}[!tbp]
\centering
\begin{minipage}{0.4\textwidth}
\begin{center}
\begin{tikzpicture}
        \tikzstyle{every state}=[
            draw = black,
            thick,
            fill = white,
            minimum size = 4mm
        ]
        
  \node[state] (Q0) at (0.1,3) {Q0};
  \node[state] (Q1) at (2.9,3)  {Q1};
  \node[state] (Q2) at (1.5,1.5)  {Q2};
  \node[state] (Q3) at (2.9,0) {Q3};
  \node[state] (Q4) at (0.1,0)  {Q4};
  
   	\draw[vecArrow] (Q0) to (Q1);
	\draw[vecArrow] (Q0) to (Q2);
	\draw[vecArrow] (Q1) to (Q2);
	\draw[vecArrow] (Q4) to (Q2);
	\draw[vecArrow] (Q4) to (Q3);
	\draw[vecArrow] (Q3) to (Q2);
\end{tikzpicture}

QX2~\cite{IBMQ2_info}
\end{center}
\end{minipage}%
\begin{minipage}{0.4\textwidth}
\begin{center}
\begin{tikzpicture}
        \tikzstyle{every state}=[
            draw = black,
            thick,
            fill = white,
            minimum size = 4mm
        ]
        
  \node[state] (Q0) at (0.1,3) {Q0};
  \node[state] (Q1) at (2.9,3)  {Q1};
  \node[state] (Q2) at (1.5,1.5)  {Q2};
  \node[state] (Q3) at (2.9,0) {Q3};
  \node[state] (Q4) at (0.1,0)  {Q4};
  
   	\draw[vecArrow] (Q3) to (Q4);
	\draw[vecArrow] (Q3) to (Q2);
	\draw[vecArrow] (Q2) to (Q4);
	\draw[vecArrow] (Q2) to (Q0);
	\draw[vecArrow] (Q2) to (Q1);
	\draw[vecArrow] (Q1) to (Q0);
\end{tikzpicture}

QX4~\cite{IBMQ4_info}
\end{center}
\end{minipage}

  \caption{Architectures of  IBM's five-qubit  processors.}
  \label{IBM_arch_fig}

\end{figure}
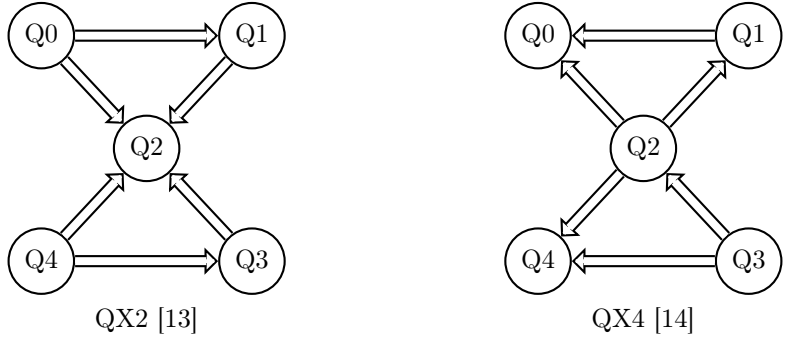

IBM has made several quantum computers available via cloud services. 
Among them are two 5-qubit quantum computers (QX2 and QX4)~\cite{IBMQ} whose 
architectures are shown in Fig.~\ref{IBM_arch_fig}.
The arrows between the qubits indicate the CNOT gates that are allowed to be implemented directly in a particular architecture. Specifically, the head of an arrow indicates the target  qubit and the other side corresponds to a control qubit. Thus, IBM quantum computers do not allow us to directly implement all the CNOT gates. However, all single qubit gates from the Clifford+T gate library can be implemented directly. Consequently, an arbitrary Clifford+T circuit may not be implementable directly in an IBM quantum processor. 
A CNOT gate that is not supported by the architecture can be realized with a sequence of gates 
(for details see~\cite{dueck_DSD_2018}). 

The Basic idea in~\cite{dueck_DSD_2018} is to find a realization for every possible CNOT operation in a given architecture.
This can be accomplished by swapping qubits until they are connected by a CNOT.
Next, the CNOT is applied and finally the qubits are swapped back to their original place.
There are many sequence of swaps that can accomplish the same objective.
For each transformation, there are also some reduction in the number of gates.
For example, to realized CNOT($Q_1,Q_4$) on QX2, the qubits $Q_2$ and $Q_4$ can be swapped, followed by 
CNOT($Q_2,Q_4$), and finally $Q_2$ and $Q_4$ are swapped again.
The SWAPs can be realized with a cost of 10 additional gates, after some reduction as shown below.


An alternative transformation with one fewer gate can be achieved as follows~\cite{dueck_DSD_2018}.
\vspace{3mm}
\begin{center}
\resizebox{5.0in}{!}{\begin{tikzpicture}[scale=1.000000,x=1pt,y=1pt]
\filldraw[color=white] (0.000000, -7.500000) rectangle (561.000000, 37.500000);
\draw[color=black] (0.000000,30.000000) -- (561.000000,30.000000);
\draw[color=black] (0.000000,30.000000) node[left] {$Q_1$};
\draw[color=black] (0.000000,15.000000) -- (561.000000,15.000000);
\draw[color=black] (0.000000,15.000000) node[left] {$Q_2$};
\draw[color=black] (0.000000,0.000000) -- (561.000000,0.000000);
\draw[color=black] (0.000000,0.000000) node[left] {$Q_3$};
\draw (9.000000,30.000000) -- (9.000000,0.000000);
\filldraw (9.000000, 30.000000) circle(1.500000pt);
\begin{scope}
\draw[fill=white] (9.000000, 0.000000) circle(3.000000pt);
\clip (9.000000, 0.000000) circle(3.000000pt);
\draw (6.000000, 0.000000) -- (12.000000, 0.000000);
\draw (9.000000, -3.000000) -- (9.000000, 3.000000);
\end{scope}
\draw[fill=white,color=white] (24.000000, -6.000000) rectangle (39.000000, 36.000000);
\draw (31.500000, 15.000000) node {$=$};
\draw (54.000000,15.000000) -- (54.000000,0.000000);
\begin{scope}
\draw (51.878680, 12.878680) -- (56.121320, 17.121320);
\draw (51.878680, 17.121320) -- (56.121320, 12.878680);
\end{scope}
\begin{scope}
\draw (51.878680, -2.121320) -- (56.121320, 2.121320);
\draw (51.878680, 2.121320) -- (56.121320, -2.121320);
\end{scope}
\draw (72.000000,30.000000) -- (72.000000,15.000000);
\filldraw (72.000000, 30.000000) circle(1.500000pt);
\begin{scope}
\draw[fill=white] (72.000000, 15.000000) circle(3.000000pt);
\clip (72.000000, 15.000000) circle(3.000000pt);
\draw (69.000000, 15.000000) -- (75.000000, 15.000000);
\draw (72.000000, 12.000000) -- (72.000000, 18.000000);
\end{scope}
\draw (90.000000,15.000000) -- (90.000000,0.000000);
\begin{scope}
\draw (87.878680, 12.878680) -- (92.121320, 17.121320);
\draw (87.878680, 17.121320) -- (92.121320, 12.878680);
\end{scope}
\begin{scope}
\draw (87.878680, -2.121320) -- (92.121320, 2.121320);
\draw (87.878680, 2.121320) -- (92.121320, -2.121320);
\end{scope}
\draw[fill=white,color=white] (105.000000, -6.000000) rectangle (120.000000, 36.000000);
\draw (112.500000, 15.000000) node {$=$};
\draw (135.000000,15.000000) -- (135.000000,0.000000);
\filldraw (135.000000, 0.000000) circle(1.500000pt);
\begin{scope}
\draw[fill=white] (135.000000, 15.000000) circle(3.000000pt);
\clip (135.000000, 15.000000) circle(3.000000pt);
\draw (132.000000, 15.000000) -- (138.000000, 15.000000);
\draw (135.000000, 12.000000) -- (135.000000, 18.000000);
\end{scope}
\begin{scope}
\draw[fill=white] (156.000000, -0.000000) +(-45.000000:8.485281pt and 8.485281pt) -- +(45.000000:8.485281pt and 8.485281pt) -- +(135.000000:8.485281pt and 8.485281pt) -- +(225.000000:8.485281pt and 8.485281pt) -- cycle;
\clip (156.000000, -0.000000) +(-45.000000:8.485281pt and 8.485281pt) -- +(45.000000:8.485281pt and 8.485281pt) -- +(135.000000:8.485281pt and 8.485281pt) -- +(225.000000:8.485281pt and 8.485281pt) -- cycle;
\draw (156.000000, -0.000000) node {$H$};
\end{scope}
\begin{scope}
\draw[fill=white] (156.000000, 15.000000) +(-45.000000:8.485281pt and 8.485281pt) -- +(45.000000:8.485281pt and 8.485281pt) -- +(135.000000:8.485281pt and 8.485281pt) -- +(225.000000:8.485281pt and 8.485281pt) -- cycle;
\clip (156.000000, 15.000000) +(-45.000000:8.485281pt and 8.485281pt) -- +(45.000000:8.485281pt and 8.485281pt) -- +(135.000000:8.485281pt and 8.485281pt) -- +(225.000000:8.485281pt and 8.485281pt) -- cycle;
\draw (156.000000, 15.000000) node {$H$};
\end{scope}
\draw (177.000000,15.000000) -- (177.000000,0.000000);
\filldraw (177.000000, 0.000000) circle(1.500000pt);
\begin{scope}
\draw[fill=white] (177.000000, 15.000000) circle(3.000000pt);
\clip (177.000000, 15.000000) circle(3.000000pt);
\draw (174.000000, 15.000000) -- (180.000000, 15.000000);
\draw (177.000000, 12.000000) -- (177.000000, 18.000000);
\end{scope}
\begin{scope}
\draw[fill=white] (198.000000, -0.000000) +(-45.000000:8.485281pt and 8.485281pt) -- +(45.000000:8.485281pt and 8.485281pt) -- +(135.000000:8.485281pt and 8.485281pt) -- +(225.000000:8.485281pt and 8.485281pt) -- cycle;
\clip (198.000000, -0.000000) +(-45.000000:8.485281pt and 8.485281pt) -- +(45.000000:8.485281pt and 8.485281pt) -- +(135.000000:8.485281pt and 8.485281pt) -- +(225.000000:8.485281pt and 8.485281pt) -- cycle;
\draw (198.000000, -0.000000) node {$H$};
\end{scope}
\begin{scope}
\draw[fill=white] (198.000000, 15.000000) +(-45.000000:8.485281pt and 8.485281pt) -- +(45.000000:8.485281pt and 8.485281pt) -- +(135.000000:8.485281pt and 8.485281pt) -- +(225.000000:8.485281pt and 8.485281pt) -- cycle;
\clip (198.000000, 15.000000) +(-45.000000:8.485281pt and 8.485281pt) -- +(45.000000:8.485281pt and 8.485281pt) -- +(135.000000:8.485281pt and 8.485281pt) -- +(225.000000:8.485281pt and 8.485281pt) -- cycle;
\draw (198.000000, 15.000000) node {$H$};
\end{scope}
\draw (219.000000,15.000000) -- (219.000000,0.000000);
\filldraw (219.000000, 0.000000) circle(1.500000pt);
\begin{scope}
\draw[fill=white] (219.000000, 15.000000) circle(3.000000pt);
\clip (219.000000, 15.000000) circle(3.000000pt);
\draw (216.000000, 15.000000) -- (222.000000, 15.000000);
\draw (219.000000, 12.000000) -- (219.000000, 18.000000);
\end{scope}
\draw (237.000000,30.000000) -- (237.000000,15.000000);
\filldraw (237.000000, 30.000000) circle(1.500000pt);
\begin{scope}
\draw[fill=white] (237.000000, 15.000000) circle(3.000000pt);
\clip (237.000000, 15.000000) circle(3.000000pt);
\draw (234.000000, 15.000000) -- (240.000000, 15.000000);
\draw (237.000000, 12.000000) -- (237.000000, 18.000000);
\end{scope}
\draw (255.000000,15.000000) -- (255.000000,0.000000);
\filldraw (255.000000, 0.000000) circle(1.500000pt);
\begin{scope}
\draw[fill=white] (255.000000, 15.000000) circle(3.000000pt);
\clip (255.000000, 15.000000) circle(3.000000pt);
\draw (252.000000, 15.000000) -- (258.000000, 15.000000);
\draw (255.000000, 12.000000) -- (255.000000, 18.000000);
\end{scope}
\begin{scope}
\draw[fill=white] (276.000000, -0.000000) +(-45.000000:8.485281pt and 8.485281pt) -- +(45.000000:8.485281pt and 8.485281pt) -- +(135.000000:8.485281pt and 8.485281pt) -- +(225.000000:8.485281pt and 8.485281pt) -- cycle;
\clip (276.000000, -0.000000) +(-45.000000:8.485281pt and 8.485281pt) -- +(45.000000:8.485281pt and 8.485281pt) -- +(135.000000:8.485281pt and 8.485281pt) -- +(225.000000:8.485281pt and 8.485281pt) -- cycle;
\draw (276.000000, -0.000000) node {$H$};
\end{scope}
\begin{scope}
\draw[fill=white] (276.000000, 15.000000) +(-45.000000:8.485281pt and 8.485281pt) -- +(45.000000:8.485281pt and 8.485281pt) -- +(135.000000:8.485281pt and 8.485281pt) -- +(225.000000:8.485281pt and 8.485281pt) -- cycle;
\clip (276.000000, 15.000000) +(-45.000000:8.485281pt and 8.485281pt) -- +(45.000000:8.485281pt and 8.485281pt) -- +(135.000000:8.485281pt and 8.485281pt) -- +(225.000000:8.485281pt and 8.485281pt) -- cycle;
\draw (276.000000, 15.000000) node {$H$};
\end{scope}
\draw (297.000000,15.000000) -- (297.000000,0.000000);
\filldraw (297.000000, 0.000000) circle(1.500000pt);
\begin{scope}
\draw[fill=white] (297.000000, 15.000000) circle(3.000000pt);
\clip (297.000000, 15.000000) circle(3.000000pt);
\draw (294.000000, 15.000000) -- (300.000000, 15.000000);
\draw (297.000000, 12.000000) -- (297.000000, 18.000000);
\end{scope}
\begin{scope}
\draw[fill=white] (318.000000, -0.000000) +(-45.000000:8.485281pt and 8.485281pt) -- +(45.000000:8.485281pt and 8.485281pt) -- +(135.000000:8.485281pt and 8.485281pt) -- +(225.000000:8.485281pt and 8.485281pt) -- cycle;
\clip (318.000000, -0.000000) +(-45.000000:8.485281pt and 8.485281pt) -- +(45.000000:8.485281pt and 8.485281pt) -- +(135.000000:8.485281pt and 8.485281pt) -- +(225.000000:8.485281pt and 8.485281pt) -- cycle;
\draw (318.000000, -0.000000) node {$H$};
\end{scope}
\begin{scope}
\draw[fill=white] (318.000000, 15.000000) +(-45.000000:8.485281pt and 8.485281pt) -- +(45.000000:8.485281pt and 8.485281pt) -- +(135.000000:8.485281pt and 8.485281pt) -- +(225.000000:8.485281pt and 8.485281pt) -- cycle;
\clip (318.000000, 15.000000) +(-45.000000:8.485281pt and 8.485281pt) -- +(45.000000:8.485281pt and 8.485281pt) -- +(135.000000:8.485281pt and 8.485281pt) -- +(225.000000:8.485281pt and 8.485281pt) -- cycle;
\draw (318.000000, 15.000000) node {$H$};
\end{scope}
\draw (339.000000,15.000000) -- (339.000000,0.000000);
\filldraw (339.000000, 0.000000) circle(1.500000pt);
\begin{scope}
\draw[fill=white] (339.000000, 15.000000) circle(3.000000pt);
\clip (339.000000, 15.000000) circle(3.000000pt);
\draw (336.000000, 15.000000) -- (342.000000, 15.000000);
\draw (339.000000, 12.000000) -- (339.000000, 18.000000);
\end{scope}
\draw[fill=white,color=white] (354.000000, -6.000000) rectangle (369.000000, 36.000000);
\draw (361.500000, 15.000000) node {$=$};
\draw (384.000000,15.000000) -- (384.000000,0.000000);
\filldraw (384.000000, 0.000000) circle(1.500000pt);
\begin{scope}
\draw[fill=white] (384.000000, 15.000000) circle(3.000000pt);
\clip (384.000000, 15.000000) circle(3.000000pt);
\draw (381.000000, 15.000000) -- (387.000000, 15.000000);
\draw (384.000000, 12.000000) -- (384.000000, 18.000000);
\end{scope}
\begin{scope}
\draw[fill=white] (405.000000, -0.000000) +(-45.000000:8.485281pt and 8.485281pt) -- +(45.000000:8.485281pt and 8.485281pt) -- +(135.000000:8.485281pt and 8.485281pt) -- +(225.000000:8.485281pt and 8.485281pt) -- cycle;
\clip (405.000000, -0.000000) +(-45.000000:8.485281pt and 8.485281pt) -- +(45.000000:8.485281pt and 8.485281pt) -- +(135.000000:8.485281pt and 8.485281pt) -- +(225.000000:8.485281pt and 8.485281pt) -- cycle;
\draw (405.000000, -0.000000) node {$H$};
\end{scope}
\begin{scope}
\draw[fill=white] (405.000000, 15.000000) +(-45.000000:8.485281pt and 8.485281pt) -- +(45.000000:8.485281pt and 8.485281pt) -- +(135.000000:8.485281pt and 8.485281pt) -- +(225.000000:8.485281pt and 8.485281pt) -- cycle;
\clip (405.000000, 15.000000) +(-45.000000:8.485281pt and 8.485281pt) -- +(45.000000:8.485281pt and 8.485281pt) -- +(135.000000:8.485281pt and 8.485281pt) -- +(225.000000:8.485281pt and 8.485281pt) -- cycle;
\draw (405.000000, 15.000000) node {$H$};
\end{scope}
\draw (426.000000,15.000000) -- (426.000000,0.000000);
\filldraw (426.000000, 0.000000) circle(1.500000pt);
\begin{scope}
\draw[fill=white] (426.000000, 15.000000) circle(3.000000pt);
\clip (426.000000, 15.000000) circle(3.000000pt);
\draw (423.000000, 15.000000) -- (429.000000, 15.000000);
\draw (426.000000, 12.000000) -- (426.000000, 18.000000);
\end{scope}
\begin{scope}
\draw[fill=white] (447.000000, 15.000000) +(-45.000000:8.485281pt and 8.485281pt) -- +(45.000000:8.485281pt and 8.485281pt) -- +(135.000000:8.485281pt and 8.485281pt) -- +(225.000000:8.485281pt and 8.485281pt) -- cycle;
\clip (447.000000, 15.000000) +(-45.000000:8.485281pt and 8.485281pt) -- +(45.000000:8.485281pt and 8.485281pt) -- +(135.000000:8.485281pt and 8.485281pt) -- +(225.000000:8.485281pt and 8.485281pt) -- cycle;
\draw (447.000000, 15.000000) node {$H$};
\end{scope}
\draw (468.000000,30.000000) -- (468.000000,15.000000);
\filldraw (468.000000, 30.000000) circle(1.500000pt);
\begin{scope}
\draw[fill=white] (468.000000, 15.000000) circle(3.000000pt);
\clip (468.000000, 15.000000) circle(3.000000pt);
\draw (465.000000, 15.000000) -- (471.000000, 15.000000);
\draw (468.000000, 12.000000) -- (468.000000, 18.000000);
\end{scope}
\begin{scope}
\draw[fill=white] (489.000000, 15.000000) +(-45.000000:8.485281pt and 8.485281pt) -- +(45.000000:8.485281pt and 8.485281pt) -- +(135.000000:8.485281pt and 8.485281pt) -- +(225.000000:8.485281pt and 8.485281pt) -- cycle;
\clip (489.000000, 15.000000) +(-45.000000:8.485281pt and 8.485281pt) -- +(45.000000:8.485281pt and 8.485281pt) -- +(135.000000:8.485281pt and 8.485281pt) -- +(225.000000:8.485281pt and 8.485281pt) -- cycle;
\draw (489.000000, 15.000000) node {$H$};
\end{scope}
\draw (510.000000,15.000000) -- (510.000000,0.000000);
\filldraw (510.000000, 0.000000) circle(1.500000pt);
\begin{scope}
\draw[fill=white] (510.000000, 15.000000) circle(3.000000pt);
\clip (510.000000, 15.000000) circle(3.000000pt);
\draw (507.000000, 15.000000) -- (513.000000, 15.000000);
\draw (510.000000, 12.000000) -- (510.000000, 18.000000);
\end{scope}
\begin{scope}
\draw[fill=white] (531.000000, -0.000000) +(-45.000000:8.485281pt and 8.485281pt) -- +(45.000000:8.485281pt and 8.485281pt) -- +(135.000000:8.485281pt and 8.485281pt) -- +(225.000000:8.485281pt and 8.485281pt) -- cycle;
\clip (531.000000, -0.000000) +(-45.000000:8.485281pt and 8.485281pt) -- +(45.000000:8.485281pt and 8.485281pt) -- +(135.000000:8.485281pt and 8.485281pt) -- +(225.000000:8.485281pt and 8.485281pt) -- cycle;
\draw (531.000000, -0.000000) node {$H$};
\end{scope}
\begin{scope}
\draw[fill=white] (531.000000, 15.000000) +(-45.000000:8.485281pt and 8.485281pt) -- +(45.000000:8.485281pt and 8.485281pt) -- +(135.000000:8.485281pt and 8.485281pt) -- +(225.000000:8.485281pt and 8.485281pt) -- cycle;
\clip (531.000000, 15.000000) +(-45.000000:8.485281pt and 8.485281pt) -- +(45.000000:8.485281pt and 8.485281pt) -- +(135.000000:8.485281pt and 8.485281pt) -- +(225.000000:8.485281pt and 8.485281pt) -- cycle;
\draw (531.000000, 15.000000) node {$H$};
\end{scope}
\draw (552.000000,15.000000) -- (552.000000,0.000000);
\filldraw (552.000000, 0.000000) circle(1.500000pt);
\begin{scope}
\draw[fill=white] (552.000000, 15.000000) circle(3.000000pt);
\clip (552.000000, 15.000000) circle(3.000000pt);
\draw (549.000000, 15.000000) -- (555.000000, 15.000000);
\draw (552.000000, 12.000000) -- (552.000000, 18.000000);
\end{scope}
\end{tikzpicture}
}	
\end{center}

\vspace{3mm}
\begin{center}
\resizebox{3.0in}{!}{\begin{tikzpicture}[scale=1.000000,x=1pt,y=1pt]
\filldraw[color=white] (0.000000, -7.500000) rectangle (312.000000, 37.500000);
\draw[color=black] (0.000000,30.000000) -- (312.000000,30.000000);
\draw[color=black] (0.000000,30.000000) node[left] {$q_1$};
\draw[color=black] (0.000000,15.000000) -- (312.000000,15.000000);
\draw[color=black] (0.000000,15.000000) node[left] {$Q_2$};
\draw[color=black] (0.000000,0.000000) -- (312.000000,0.000000);
\draw[color=black] (0.000000,0.000000) node[left] {$Q_3$};
\draw (9.000000,30.000000) -- (9.000000,0.000000);
\filldraw (9.000000, 30.000000) circle(1.500000pt);
\begin{scope}
\draw[fill=white] (9.000000, 0.000000) circle(3.000000pt);
\clip (9.000000, 0.000000) circle(3.000000pt);
\draw (6.000000, 0.000000) -- (12.000000, 0.000000);
\draw (9.000000, -3.000000) -- (9.000000, 3.000000);
\end{scope}
\draw[fill=white,color=white] (24.000000, -6.000000) rectangle (39.000000, 36.000000);
\draw (31.500000, 15.000000) node {$=$};
\draw (54.000000,30.000000) -- (54.000000,15.000000);
\filldraw (54.000000, 30.000000) circle(1.500000pt);
\begin{scope}
\draw[fill=white] (54.000000, 15.000000) circle(3.000000pt);
\clip (54.000000, 15.000000) circle(3.000000pt);
\draw (51.000000, 15.000000) -- (57.000000, 15.000000);
\draw (54.000000, 12.000000) -- (54.000000, 18.000000);
\end{scope}
\draw (72.000000,15.000000) -- (72.000000,0.000000);
\filldraw (72.000000, 15.000000) circle(1.500000pt);
\begin{scope}
\draw[fill=white] (72.000000, 0.000000) circle(3.000000pt);
\clip (72.000000, 0.000000) circle(3.000000pt);
\draw (69.000000, 0.000000) -- (75.000000, 0.000000);
\draw (72.000000, -3.000000) -- (72.000000, 3.000000);
\end{scope}
\draw (90.000000,30.000000) -- (90.000000,15.000000);
\filldraw (90.000000, 30.000000) circle(1.500000pt);
\begin{scope}
\draw[fill=white] (90.000000, 15.000000) circle(3.000000pt);
\clip (90.000000, 15.000000) circle(3.000000pt);
\draw (87.000000, 15.000000) -- (93.000000, 15.000000);
\draw (90.000000, 12.000000) -- (90.000000, 18.000000);
\end{scope}
\draw (108.000000,15.000000) -- (108.000000,0.000000);
\filldraw (108.000000, 15.000000) circle(1.500000pt);
\begin{scope}
\draw[fill=white] (108.000000, 0.000000) circle(3.000000pt);
\clip (108.000000, 0.000000) circle(3.000000pt);
\draw (105.000000, 0.000000) -- (111.000000, 0.000000);
\draw (108.000000, -3.000000) -- (108.000000, 3.000000);
\end{scope}
\draw[fill=white,color=white] (123.000000, -6.000000) rectangle (138.000000, 36.000000);
\draw (130.500000, 15.000000) node {$=$};
\draw (153.000000,30.000000) -- (153.000000,15.000000);
\filldraw (153.000000, 30.000000) circle(1.500000pt);
\begin{scope}
\draw[fill=white] (153.000000, 15.000000) circle(3.000000pt);
\clip (153.000000, 15.000000) circle(3.000000pt);
\draw (150.000000, 15.000000) -- (156.000000, 15.000000);
\draw (153.000000, 12.000000) -- (153.000000, 18.000000);
\end{scope}
\begin{scope}
\draw[fill=white] (174.000000, -0.000000) +(-45.000000:8.485281pt and 8.485281pt) -- +(45.000000:8.485281pt and 8.485281pt) -- +(135.000000:8.485281pt and 8.485281pt) -- +(225.000000:8.485281pt and 8.485281pt) -- cycle;
\clip (174.000000, -0.000000) +(-45.000000:8.485281pt and 8.485281pt) -- +(45.000000:8.485281pt and 8.485281pt) -- +(135.000000:8.485281pt and 8.485281pt) -- +(225.000000:8.485281pt and 8.485281pt) -- cycle;
\draw (174.000000, -0.000000) node {$H$};
\end{scope}
\begin{scope}
\draw[fill=white] (174.000000, 15.000000) +(-45.000000:8.485281pt and 8.485281pt) -- +(45.000000:8.485281pt and 8.485281pt) -- +(135.000000:8.485281pt and 8.485281pt) -- +(225.000000:8.485281pt and 8.485281pt) -- cycle;
\clip (174.000000, 15.000000) +(-45.000000:8.485281pt and 8.485281pt) -- +(45.000000:8.485281pt and 8.485281pt) -- +(135.000000:8.485281pt and 8.485281pt) -- +(225.000000:8.485281pt and 8.485281pt) -- cycle;
\draw (174.000000, 15.000000) node {$H$};
\end{scope}
\draw (195.000000,15.000000) -- (195.000000,0.000000);
\filldraw (195.000000, 0.000000) circle(1.500000pt);
\begin{scope}
\draw[fill=white] (195.000000, 15.000000) circle(3.000000pt);
\clip (195.000000, 15.000000) circle(3.000000pt);
\draw (192.000000, 15.000000) -- (198.000000, 15.000000);
\draw (195.000000, 12.000000) -- (195.000000, 18.000000);
\end{scope}
\begin{scope}
\draw[fill=white] (216.000000, 15.000000) +(-45.000000:8.485281pt and 8.485281pt) -- +(45.000000:8.485281pt and 8.485281pt) -- +(135.000000:8.485281pt and 8.485281pt) -- +(225.000000:8.485281pt and 8.485281pt) -- cycle;
\clip (216.000000, 15.000000) +(-45.000000:8.485281pt and 8.485281pt) -- +(45.000000:8.485281pt and 8.485281pt) -- +(135.000000:8.485281pt and 8.485281pt) -- +(225.000000:8.485281pt and 8.485281pt) -- cycle;
\draw (216.000000, 15.000000) node {$H$};
\end{scope}
\draw (237.000000,30.000000) -- (237.000000,15.000000);
\filldraw (237.000000, 30.000000) circle(1.500000pt);
\begin{scope}
\draw[fill=white] (237.000000, 15.000000) circle(3.000000pt);
\clip (237.000000, 15.000000) circle(3.000000pt);
\draw (234.000000, 15.000000) -- (240.000000, 15.000000);
\draw (237.000000, 12.000000) -- (237.000000, 18.000000);
\end{scope}
\begin{scope}
\draw[fill=white] (258.000000, 15.000000) +(-45.000000:8.485281pt and 8.485281pt) -- +(45.000000:8.485281pt and 8.485281pt) -- +(135.000000:8.485281pt and 8.485281pt) -- +(225.000000:8.485281pt and 8.485281pt) -- cycle;
\clip (258.000000, 15.000000) +(-45.000000:8.485281pt and 8.485281pt) -- +(45.000000:8.485281pt and 8.485281pt) -- +(135.000000:8.485281pt and 8.485281pt) -- +(225.000000:8.485281pt and 8.485281pt) -- cycle;
\draw (258.000000, 15.000000) node {$H$};
\end{scope}
\draw (279.000000,15.000000) -- (279.000000,0.000000);
\filldraw (279.000000, 0.000000) circle(1.500000pt);
\begin{scope}
\draw[fill=white] (279.000000, 15.000000) circle(3.000000pt);
\clip (279.000000, 15.000000) circle(3.000000pt);
\draw (276.000000, 15.000000) -- (282.000000, 15.000000);
\draw (279.000000, 12.000000) -- (279.000000, 18.000000);
\end{scope}
\begin{scope}
\draw[fill=white] (300.000000, -0.000000) +(-45.000000:8.485281pt and 8.485281pt) -- +(45.000000:8.485281pt and 8.485281pt) -- +(135.000000:8.485281pt and 8.485281pt) -- +(225.000000:8.485281pt and 8.485281pt) -- cycle;
\clip (300.000000, -0.000000) +(-45.000000:8.485281pt and 8.485281pt) -- +(45.000000:8.485281pt and 8.485281pt) -- +(135.000000:8.485281pt and 8.485281pt) -- +(225.000000:8.485281pt and 8.485281pt) -- cycle;
\draw (300.000000, -0.000000) node {$H$};
\end{scope}
\begin{scope}
\draw[fill=white] (300.000000, 15.000000) +(-45.000000:8.485281pt and 8.485281pt) -- +(45.000000:8.485281pt and 8.485281pt) -- +(135.000000:8.485281pt and 8.485281pt) -- +(225.000000:8.485281pt and 8.485281pt) -- cycle;
\clip (300.000000, 15.000000) +(-45.000000:8.485281pt and 8.485281pt) -- +(45.000000:8.485281pt and 8.485281pt) -- +(135.000000:8.485281pt and 8.485281pt) -- +(225.000000:8.485281pt and 8.485281pt) -- cycle;
\draw (300.000000, 15.000000) node {$H$};
\end{scope}
\end{tikzpicture}
}	
\end{center}
The algorithmic approach adopted here requires that all such transformations, with their corresponding cost (number of additional gates) are to be stored in a table.
This table has to be build only once for every architecture.
It will take one table look up to find the best realization for CNOT($a,b$).
All CNOT gates in a given circuit Clifford+T circuits are to be replaced with their equivalent transformations.

It is clear that different mappings of logical qubits to the physical ones will yield different costs in the target architecture.
For example, if a circuit contains the following gates CNOT($a,b$) and CNOT($b,c$). 
An optimal mapping for QX2 would be $\{ a \rightarrow Q0, b \rightarrow Q1, c \rightarrow Q2 \}$ whereas for
QX4 it would be $\{ a \rightarrow Q3, b \rightarrow Q2, c \rightarrow Q0 \}$. Thus, the optimized circuit to be obtained would depend on the architecture of the quantum processor, but the method to be followed is independent of the architecture.

For a circuit with only five qubits, it is feasible to calculate the cost for all 120 permutations and pick the best one.
this is exactly what is proposed in~\cite{dueck_DSD_2018}.
Finally, some simplifications may be possible.
For example, the transformation may yield two consecutive Hadamard gates on the same qubits, which can be removed. Other more complex circuit identities or templates can also be used.


%

\section{Benefit of the optimization: An illustrative examples\label{sec:Benefit-of-the-optimization}}
We have already mentioned that various quantum computing tasks have been performed using IBM's QX2 and QX4, but the corresponding circuits were not optimized. In this section, we select some of those circuits to establish that the optimized circuit can yield better results and reveal quantumness in a stronger manner. Specifically, we have chosen a set of circuits implemented in QX2  to demonstrate  experimental violation of Mermin inequalities. These circuits are of particular importance for various reasons, especially for the fact that Mermin inequalities being an extension of Bell's inequalities to the multi-partite scenario can be used to discriminate between classical physics and quantum physics, and establish the nonlocal nature of the physical world. In other words, an experimentally observed violation of a Mermin inequality can strongly establish that the physical world cannot be described by local hidden variable theories. This is somewhat obvious as Mermin inequities are essentially Bell type inequalities. Violation of Bell's inequality has been shown in many experiments, but it's slightly more difficult to show the violation of Mermin inequality as it's difficult to achieve a good control of three or more qubits, including the
generation of entangled states (say a GHZ-type state which maximally violates Mermin inequality) \cite{AL2016experimental}. This is where an optimized circuit may play a crucial role by providing greater control and higher fidelity.

\subsection{Circuits for the realization of Mermin inequality}

\begin{figure}[h]

\includegraphics[scale=0.6]{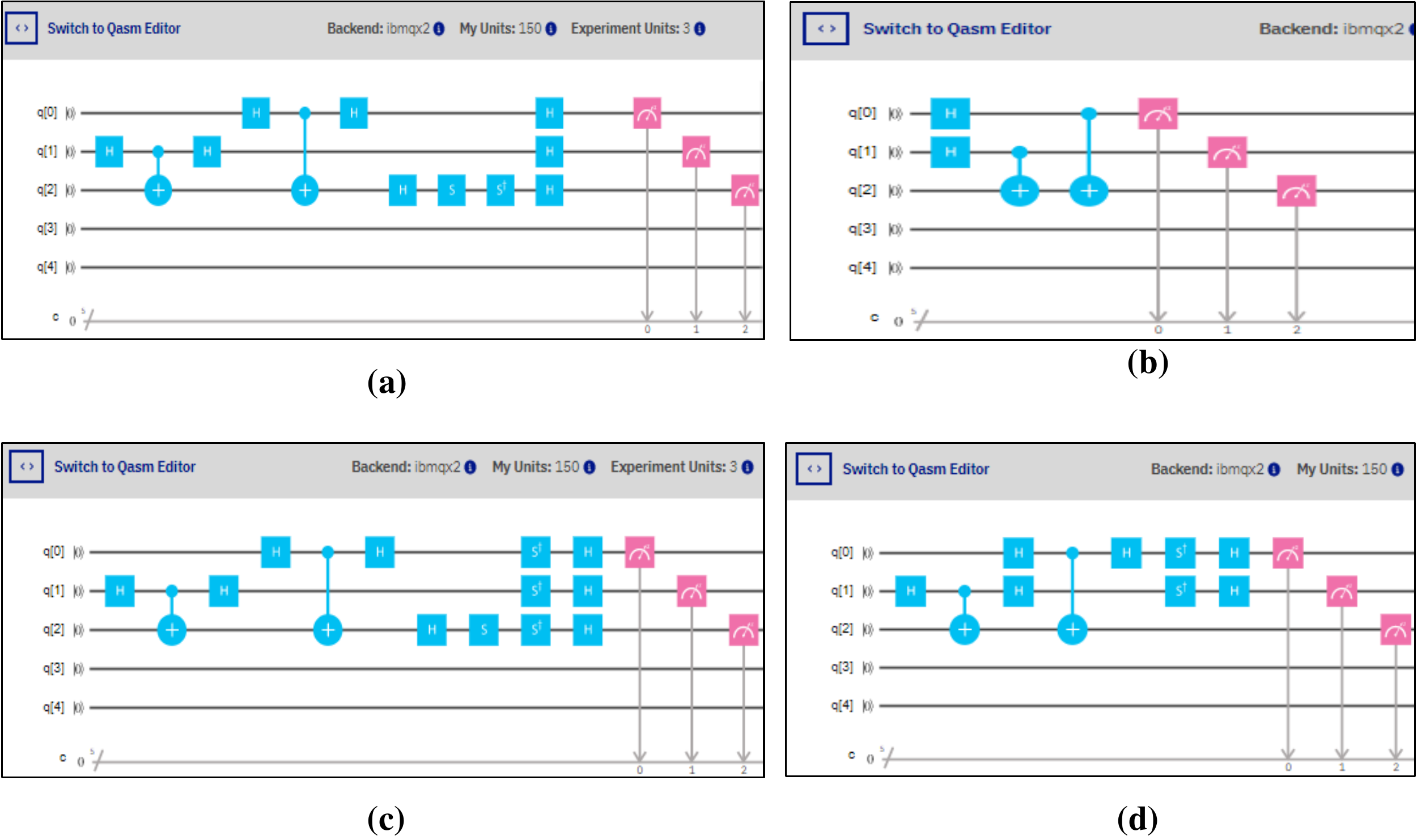}\caption{\label{fig:fig1}(Color online) (a) and (c) show the circuits used in Fig.\ 1 of~\cite{AL2016experimental} to demonstrate violation of Mermin inequality in tripartite scenario. Optimized equivalent circuits for (a) and (c) are shown as  (b) and (d), respectively.} 
\label{Mermin-circuits}
\end{figure}

Violation of any of the Mermin inequalities implies the existence of a situation that cannot  be explained classically. In the recent past, violation of Mermin inequality has been shown using various techniques, including ion-trap~\cite{lanyon2014experimental}, and optics-based~\cite{zhao2003experimental} techniques. One of the most recent experiment in this line has been done by Alsina and Latorre in 2016~\cite{AL2016experimental}. They have demonstrated violation of Mermin inequalities using IBM's QX2 quantum computing processor. 
The work was done almost immediately after the access to QX2 was provided. In their work, they had shown violation of Mermin inequalities for $n=3$ to $n=5$. For $n=3$ case, i.e., in a tripartite scenario, they used the circuits shown in Fig.\ \ref{Mermin-circuits} (a) and (c) (cf. Fig.\ 1 of~\cite{AL2016experimental}). We have used our algorithm to optimize these circuits. The optimized circuits corresponding to the circuits shown in  Fig.\ \ref{Mermin-circuits} (a) and (c)  are shown in Fig.\ \ref{Mermin-circuits} (b) and (d),  respectively.
It is clearly visible that the optimized circuits have smaller gate counts and levels. 
Specifically the circuit shown in Fig.\ \ref{Mermin-circuits} (a) has 12 gates and 7 levels, but the corresponding optimized circuit (shown in Fig.\ \ref{Mermin-circuits} (b)) has only 4 gates and 3 levels. 
Thus, the gate count has been reduced by 67\% and the number of levels have been reduced by 57\%. 
Similarly, for the circuit shown in Fig.\ \ref{Mermin-circuits} (c) has 14 gates and 8 levels, but the corresponding optimized circuit (as shown in  Fig.\ \ref{Mermin-circuits} (d)) has only  10 gates and 6 levels. 
Thus, the optimization scheme adopted here can considerably reduce the circuit costs and provide efficient circuits. However, reduction in circuit costs is not the only parameter that can quantitatively illustrate the necessity and advantages of the optimization procedure. There are other ways to check whether the optimization process has improved the performance of the circuit.

\begin{figure}

\begin{centering}
\includegraphics[scale=0.6]{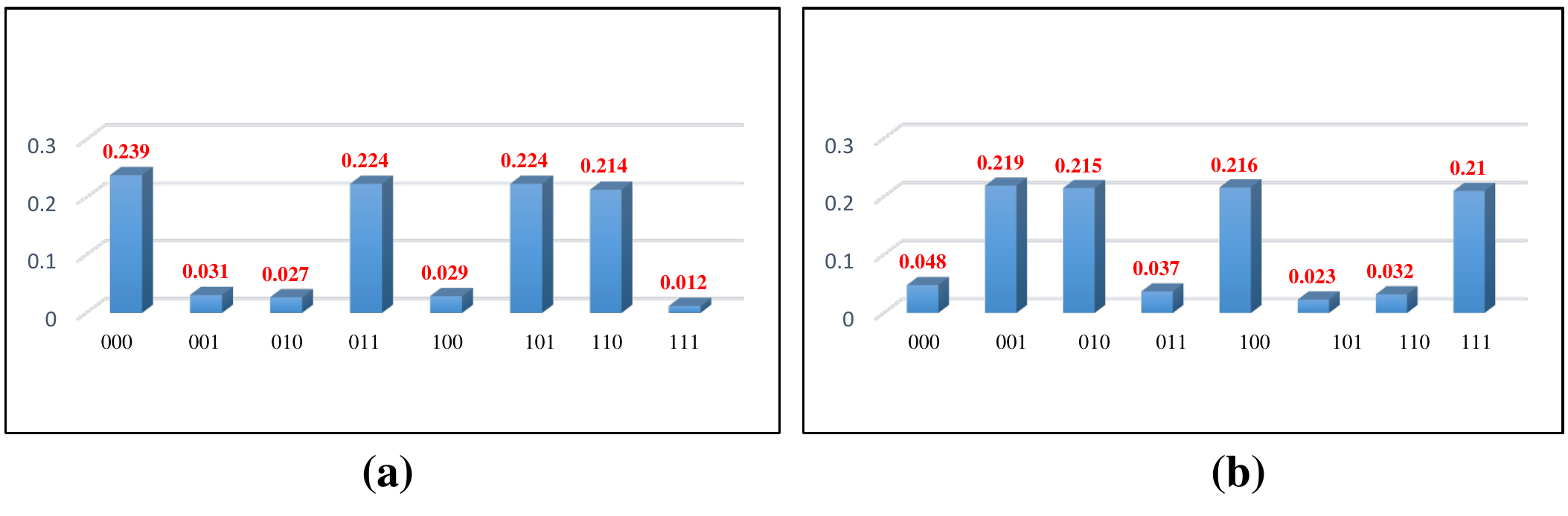}\caption{\label{fig:fig2}(Color online) ((a) and (b)) represent the obtained probability distribution for circuits shown in Figs. \ref{fig:fig1} (a) and (b), respectively.}
\par\end{centering}
\end{figure}

To begin with we run the circuits shown in Fig. \ref{Mermin-circuits} for 8,192 times each by using QX2 processor. Originally, in~\cite{AL2016experimental}, the unoptimized circuits were run for 1,024 times. The output of our experiment is specifically shown in Fig.\ \ref{fig:fig2}, where the left panel corresponds to the output of the circuit shown in Fig. \ref{Mermin-circuits} (a) whereas the right panel illustrates the experimental output of the corresponding optimized circuit. These outputs are just representative. The complete set of similar results obtained by realizing the circuits shown in Fig.\ \ref{Mermin-circuits}  using QX2 is given in Table~\ref{Mermin-experiment}. This table can be used to compute violation of Mermin inequality in each case. Here, we may note that for $n=3$, i.e., for 3-qubit case, the expectation value of the Mermin polynomial for a classical theory (which essentially obeys local realism) is bounded by 2 (i.e., $\langle M_3\rangle^{\rm{classical}} \leq 2$), whereas for quantum mechanics (QM) it is bounded by 4 (i.e., $\langle 
 M_3\rangle^{\rm{QM}} \leq 4$).  Now, we may consider $\langle M_3\rangle-2$ as a measure of how strongly Mermin inequality is violated by a 3-qubit state. In fact, we can calculate $\langle M_{3} \rangle$ using Table~\ref{Mermin-experiment}, by  following the method used in \cite{AL2016experimental}. This method uses  the formula $\langle M_3\rangle= 3\langle XXY \rangle - \langle YYY \rangle$ where  $\langle A \rangle = \sum_{i} \rm{P}_{i}E_{i}$, with $\rm{P}_{i}$ is the probability of system being found in the $i^{th}$ eigenstate of operator $\operatorname{A}$ and $E_{i}$ is the eigenvalue of the corresponding state. The experimental setup for the measurement of $\langle XXY \rangle (\langle YYY \rangle)$ is given in Figs.\ \ref{fig:fig1} a and c (Figs.\ \ref{fig:fig1} b and d). Thus, using $\rm{P}_{i}$s given in Table \ref{Mermin-experiment} we can compute $\langle M_3\rangle$ which is  3.126 in our case in contrast to the value of $2.85\pm0.02$ obtained in the original work of Alsina and Latorre \cite{AL2016experimental}. As we have already mentioned that  $\langle M_3\rangle>2$  implies the violation of Mermin inequality or the nonexistence of a classical local realistic (LR) theory, our result establishes the existence of a nonclassical theory in general (quantum mechanics in particular). The observed value of 3.126 (>2.85) for optimized case indicates that the optimized circuits can witness the signature of nonclassicality in a stronger manner. Thus, in those cases where Marmin inequalities or other similar inequalities are violated weakly,  optimized circuits will be of much relevance in identifying the signature of nonclassicality.

\begin{table}
\centering
\begin{tabular}{|c|c|c|c|c|c|c|c|c|c|}
\hline 
Circuit & No.\ experiments  & \rm{P$_{000}$} & \rm{P$_{001}$} & \rm{P$_{010}$} & \rm{P$_{011}$} & \rm{P$_{100}$} & \rm{P$_{101}$} & \rm{P$_{110}$} & \rm{P$_{111}$} \\ 
\hline 
Fig.\ \ref{fig:fig1} a  & 1,024  & 0.229 & 0.042 & 0.024 & 0.194 & 0.043 & 0.203 & 0.231 & 0.033 \tabularnewline
\hline 
Fig.\ \ref{fig:fig1} a  & 8,192  & 0.238 & 0.041 & 0.035 & 0.202 & 0.031 & 0.223 & 0.217 & 0.013\tabularnewline
\hline 
Fig.\ \ref{fig:fig1} b & 8,192 & 0.239 & 0.031 & 0.027 & 0.224 & 0.029 & 0.224 & 0.214 & 0.012\tabularnewline
\hline 
 Fig.\ \ref{fig:fig1} c  & 1,024   & 0.050 & 0.188 & 0.188 & 0.028 & 0.258 & 0.026 & 0.041 & 0.221\tabularnewline
\hline 
Fig.\ \ref{fig:fig1} c & 8,192   & 0.046 & 0.223 & 0.210 & 0.033 & 0.218 & 0.029 & 0.028 & 0.214\tabularnewline
\hline 
Fig.\ \ref{fig:fig1} d & 8,192 & 0.048 & 0.219 & 0.215 & 0.037 & 0.216 & 0.023 & 0.032 & 0.210\tabularnewline
\hline 
\end{tabular}
\caption{Results of Mermin experiment are provided in the table. $\rm{P_{i}}$ show the probability of finding the state $\rm{|i\rangle}$ on measuring the output state in the computational basis. Results of the circuit given in the Fig. \ref{fig:fig1} a are shown in the first row for 1024 runs and for 8192 runs in the second row. Results of the circuit given in Fig. \ref{fig:fig1} b  are provided in the second row for 8192 runs. Similarly, results of the Fig. \ref{fig:fig1} c are given in the rows third and fourth for 1024 and 8192 runs, respectively. Ultimately, results of the circuit in the Fig. \ref{fig:fig1} d are given in last row.} \label{Mermin-experiment}
\end{table}

\begin{table} 
\begin{tabular}{|c|c|c|c|}
\hline 
No. of qubits & \cite{AL2016experimental}'s circuits (1024 runs) & \cite{AL2016experimental}'s circuits (8192 runs) & Optimized circuits (8192 runs)\tabularnewline
\hline 
3 qubits   & $2.85\pm0.02$ & 3.009 & 3.126\tabularnewline
\hline 
\end{tabular}
\caption{A comparison of the value of $\langle M_3\rangle$ obtained originally in \cite{AL2016experimental}, where the experiment was run for
1024 times
with that obtained by running the same experiment for 8192 times and running the optimized circuit for 8192 times.} \label{tab2}
\end{table}


Even this set of experiments does not give the whole picture. To further illustrate the benefit of the circuit optimization procedure adopted here, we need to perform quantum state tomography (QST). In order to perform QST using IBM quantum processors, one can adopt the procedure described in our earlier works (see~\cite{sisodia2017experimental}) which will require some additional experiments. Here, we restrict ourselves from describing the procedure adopted for performing QST and directly report the relevant density matrices.

Ideal density matrix corresponding to the output state of the circuit shown in Fig.\ \ref{fig:fig1} (a) is
\begin{equation}
\rho_{ideal}= \left(\begin{array}{cccccccc}
\frac{1}{4} & 0 & 0 & \frac{1}{4} & 0 & \frac{1}{4} & \frac{1}{4} & 0\\
0 & 0 & 0 & 0 & 0 & 0 & 0 & 0\\
0 & 0 & 0 & 0 & 0 & 0 & 0 & 0\\
\frac{1}{4} & 0 & 0 & \frac{1}{4} & 0 & \frac{1}{4} & \frac{1}{4} & 0\\
0 & 0 & 0 & 0 & 0 & 0 & 0 & 0\\
\frac{1}{4} & 0 & 0 & \frac{1}{4} & 0 & \frac{1}{4} & \frac{1}{4} & 0\\
\frac{1}{4} & 0 & 0 & \frac{1}{4} & 0 & \frac{1}{4} & \frac{1}{4} & 0\\
0 & 0 & 0 & 0 & 0 & 0 & 0 & 0
\end{array}\right).\label{eq:id}
\end{equation} 

Density matrix of the actual state generated in the experimental realization of the circuit shown in  Fig.\ \ref{fig:fig1} (a) is obtained by QST, and the  real and imaginary parts of that density matrix is given below
\begin{equation}Re\left[\rho_{ex}A\right] =\left(\begin{array}{cccccccc}
0.238 & 0.03 & 0.007 & 0.098 & 0.005 & 0.115 & 0.095 & -0.016\\
0.03 & 0.031 & 0.094 & 0.019 & 0.118 & 0.002 & 0.012 & 0.094\\
0.007 & 0.094 & 0.035 & -0.002 & 0.091 & -0.009 & 0.001 & 0.097\\
0.098 & 0.019 & -0.002 & 0.217 & 0.017 & 0.093 & 0.097 & 0.002\\
0.005 & 0.118 & 0.091 & 0.017 & 0.041 & 0.030 & 0.034 & 0.095\\
0.115 & 0.002 & -0.009 & 0.093 & 0.030 & 0.223 & 0.094 & -0.02\\
0.095 & 0.012 & 0.001 & 0.097 & 0.034 & 0.094 & 0.202 & -0.016\\
-0.016 & 0.094 & 0.097 & 0.002 & 0.095 & -0.02 & -0.016 & 0.013
\end{array}\right)\label{eq:exA-r},
\end{equation}
and
\begin{equation}Im\left[\rho_{ex}A\right]=\left(\begin{array}{cccccccc}
0 & 0.005 & 0.002 & -0.065 & -0.002 & -0.01 & -0.029 & -0.005\\
-0.005 & 0 & -0.001 & -0.0045 & 0.009 & -0.002 & -0.003 & -0.019\\
-0.002 & 0.001 & 0 & -0.012 & 0.034 & -0.0015 & 0.002 & 0.001\\
0.065 & 0.004 & 0.012 & 0 & 0.009 & 0.021 & 0.002 & 0\\
0.002 & -0.009 & -0.034 & -0.009 & 0 & -0.006 & 0.002 & -0.0002\\
0.01 & 0.002 & 0.001 & -0.021 & 0.006 & 0.223 & -0.049 & -0.006\\
0.029 & 0.003 & -0.002 & -0.002 & -0.002 & 0.049 & 0 & 0.006\\
0.0052 & 0.019 & -0.001 & 0 & 0.0002 & 0.006 & -0.006 & 0
\end{array}\right),\label{eq:exA-i}.
\end{equation}
Similarly, real and imaginary parts of experimental density matrix of the output state of the corresponding optimized circuit shown in Fig.\ \ref{fig:fig1} (b) can be obtained by QST as 
\begin{equation}Re\left[\rho_{ex}B\right] = \left(\begin{array}{cccccccc}
0.239 & 0.032 & -0.004 & 0.199 & 0.032 & 0.21 & 0.183 & -0.014\\
0.032 & 0.029 & 0.005 & 0.021 & 0.023 & 0.023 & 0.019 & 0.007\\
-0.004 & 0.005 & 0.027 & -0.0125 & 0.004 & -0.003 & -0.009 & 0.010\\
0.199 & 0.021 & -0.012 & 0.214 & 0.035 & 0.182 & 0.197 & -0.016\\
0.032 & 0.023 & 0.004 & 0.035 & 0.031 & 0.031 & 0.030 & -0.001\\
0.21 & 0.023 & -0.003 & 0.182 & 0.031 & 0.224 & 0.184 & -0.019\\
0.18 & 0.019 & -0.009 & 0.197 & 0.030 & 0.184 & 0.224 & -0.019\\
-0.014 & 0.007 & 0.010 & -0.0165 & -0.001 & -0.019 & -0.019 & 0.012
\end{array}\right)\label{eq:exB-r}
\end{equation}
and
\begin{equation}Im\left[\rho_{ex}B\right] =\left(\begin{array}{cccccccc}
0 & -0.009 & 0.004 & -0.050 & -0.010 & 0.002 & -0.064 & -0.010\\
0.009 & 0 & 0.0002 & -0.008 & 0.001 & -\text{\ensuremath{4.3\textasciicircum-19}} & -0.007 & -0.019\\
-0.004 & -0.0002 & 0 & -0.014 & -0.010 & -0.008 & -0.016 & 0.005\\
0.050 & 0.008 & 0.014 & 0 & 0.008 & 0.032 & -0.008 & -0.007\\
0.01 & -0.001 & 0.010 & -0.0081 & 0 & -0.001 & -0.005 & 0.002\\
-0.002 & \text{\ensuremath{4.3\textasciicircum-19}} & 0.008 & -0.032 & 0.001 & 0 & -0.050 & -0.004\\
0.064 & 0.007 & 0.016 & 0.008 & 0.005 & 0.050 & 0 & -0.009\\
0.01 & 0.019 & -0.005 & 0.007 & -0.002 & 0.004 & 0.0095 & 0
\end{array}\right).\label{eq:exB-i}
\end{equation}

To establish the relevance of the present study, i.e., optimization of quantum circuits, we have used a distance based measure to quantify the performance of a quantum circuit known as fidelity, which is defined as  $F=Tr[\sqrt{\sqrt{\rho_1}.\rho_2.\sqrt{\rho_1}}]$, where $\rho_1$ and $\rho_2$ correspond to two quantum states to be compared. Specifically, we have 
computed the fidelity between the quantum state expected in the ideal scenario ($\rho_{ideal}$ in Eq. (\ref{eq:id})) with that obtained in the real experiments  using QX2 ($\rho_{ex}A$ in Eq. (\ref{eq:exA-r}) and in Eq. (\ref{eq:exA-i})). Thereafter, we have computed the fidelity of quantum state ($\rho_{ideal}$ in Eq. (\ref{eq:id})) and that obtained by the experiment performed using our optimized quantum circuit ($\rho_{ex}B$ in Eq. (\ref{eq:exB-r}) and in Eq. (\ref{eq:exB-i})). 
For example, in case of Fig.\ \ref{fig:fig1} (a), we obtained the  fidelity for the original circuit as 0.72, while that of our optimized circuit in Fig.\ \ref{fig:fig1} (b) is 0.90. Similarly, for Fig.\ \ref{fig:fig1} (c), we obtained the  fidelity of the original circuit as 0.88, while that of our optimized circuit in Fig.\ \ref{fig:fig1} (d) is obtained to be 0.89. Thus, the optimization procedure clearly helps us in performing quantum computation  with greater accuracy and to prepare the desired quantum states with higher fidelity.

%
%
%
%

\section{More results: Optimized circuits for various quantum computing tasks
\label{sec:More-results:-Optimized}}

The algorithmic approach developed in~\cite{dueck_DSD_2018} and followed here has been used to
optimize various quantum circuits. They are summarized in Tables \ref{QX4a}-\ref{QX4b}. Specifically, Tables \ref{QX4a} and \ref{QX2} report results of circuit optimization algorithm applied on circuits for various quantum computing tasks using QX4 and QX2 processors, respectively. Interested readers may access the optimized circuits along with corresponding original circuits at 
https://github.com/QBenchmark/benchmarks \cite{benchmarks}. From the Tables \ref{QX4a}-\ref{QX4b} and 
https://github.com/QBenchma\\rk/benchmarks \cite{benchmarks}, it's clear that the optimization algorithm  decreases gate counts and number of levels in most of the cases. 
In particular, we observe the best results of optimization of circuits for non-destructive discrimination of arbitrary set of orthogonal quantum states given in~\cite{majumder2018experimental} for both QX4 and QX2 architectures  (for details, see Tables \ref{QX4a} and \ref{QX2}: (67-78) \% of reduction in gate counts and (57-63)\% reduction in the levels are achieved for the circuits shown in Fig.\ (25-29) and in Fig.\ 1a of ~\cite{majumder2018experimental}). The optimization performances for various other circuits (see corresponding papers cited in the table) indexed in decreasing order are given. Table \ref{QX4b} contains optimization details of another set of circuits (which can be best described as reversible circuits) as given in~\cite{RevLib}. Here we observe a 
maximum reduction of 72\% in number of gates for circuit for evaluating the function \textbf{alu-v1\_28} and 70\% in the number of levels for circuit \textbf{alu-v4\textunderscore37}.  Clearly, the method used here can reduce the gate count and number of levels for most of the circuits implemented so far using IBM quantum processors. It can also efficiently optimize a large number of reversible circuits.

\begin{table} \begin{center}
\begin{tabular}{|c|c|c|c|c|c|c|c|}
\cline{3-8}
\multicolumn{2}{c}{\textbf{}}                &      \multicolumn{2}{|c|}{\textbf{Initial}}   &      \multicolumn{2}{|c|}{\textbf{Final}}  &      \multicolumn{2}{|c|}{\textbf{\% Reduction}} \\ \hline
\textbf{Ref} & \textbf{Fig} & \textbf{ Gates} & \textbf{ Levels} & \textbf{ Gates} & \textbf{ Levels} & \textbf{ Gates} & \textbf{ Levels} \\ \hline
\cite{majumder2018experimental} & 26  & 18 & 8  & 4  & 3  & 78 & 63  \\ \hline
\cite{majumder2018experimental} & 27  & 19 & 9  & 5  & 4  & 74 & 56  \\ \hline
\cite{majumder2018experimental} & 29  & 19 & 8  & 5  & 4  & 74 & 50  \\ \hline
\cite{majumder2018experimental} & 25  & 17 & 8  & 5  & 4  & 71 & 50  \\ \hline
\cite{alsina2016experimental}  & 1a  & 12 & 7  & 4  & 2  & 67 & 71  \\ \hline
\cite{majumder2018experimental} & 28  & 20 & 8  & 8  & 4  & 60 & 50  \\ \hline
\cite{sisodia2017design} & 3   & 28 & 22 & 13 & 11 & 54 & 50  \\ \hline
\cite{deffner2017demonstration} & 5   & 4  & 3  & 2  & 2  & 50 & 33  \\ \hline
\cite{martin2018five}  & 13a & 8  & 6  & 4  & 3  & 50 & 50  \\ \hline
\cite{alsina2016experimental}  & 1b  & 14 & 7  & 8  & 5  & 43 & 29  \\ \hline
\cite{deffner2017demonstration}  & 3   & 11 & 5  & 7  & 5  & 36 & 0   \\ \hline
\cite{sisodia2017experimental} & 4b  & 11 & 7  & 7  & 5  & 36 & 29  \\ \hline
\cite{alsina2016experimental}  & 2a  & 17 & 7  & 11 & 6  & 35 & 14  \\ \hline
\cite{deffner2017demonstration}  & 6   & 6  & 4  & 4  & 4  & 33 & 0   \\ \hline
\cite{bikash2018purification}  & 3   & 37 & 21 & 25 & 15 & 32 & 29  \\ \hline
\cite{deffner2017demonstration}  & 4   & 19 & 7  & 13 & 7  & 32 & 0   \\ \hline
\cite{srinivasan2017solving} & 2   & 13 & 9  & 9  & 7  & 31 & 22  \\ \hline
\cite{alsina2016experimental}  & 2b  & 20 & 7  & 14 & 8  & 30 & -14 \\ \hline
\cite{sisodia2017experimental}  & 8   & 28 & 19 & 22 & 17 & 21 & 11  \\ \hline
\cite{martin2018five}  & 13b & 10 & 6  & 8  & 5  & 20 & 17  \\ \hline
\cite{sisodia2017experimental}  & 3b  & 11 & 8  & 9  & 7  & 18 & 13  \\ \hline
\cite{li2017approximate}  & 6   & 41 & 27 & 34 & 20 & 17 & 26  \\ \hline
\cite{vishnu2018experimental} & 6   & 24 & 13 & 20 & 11 & 17 & 15  \\ \hline
\cite{kalra2017demonstration}  & 2   & 13 & 10 & 11 & 9  & 15 & 10  \\ \hline
\cite{vishnu2018experimental} & 4   & 16 & 10 & 14 & 7  & 13 & 30  \\ \hline
\cite{kalra2017demonstration}  & 3   & 17 & 14 & 15 & 11 & 12 & 21  \\ \hline
\cite{satyajit2018nondestructive} & 6   & 23 & 15 & 21 & 14 & 9  & 7   \\ \hline
\cite{bikash2018purification}  & 6   & 29 & 20 & 27 & 20 & 7  & 0   \\ \hline
\cite{kamal2017fixed-point} & 1   & 53 & 31 & 50 & 31 & 6  & 0   \\ \hline
\cite{ghosh2018automated}  & 8   & 57 & 38 & 55 & 37 & 4  & 3   \\ \hline
\end{tabular}
\caption{Results of the optimization of the quantum circuits which were implemented earlier using IBM quantum processor. Here, the optimization is done using our algorithm and by considering the architecture used in QX4} \label{QX4a}
\end{center} \end{table}

\begin{table} \begin{center}
\begin{tabular}{|c|c|c|c|c|c|c|c|}
\cline{3-8}
                       \multicolumn{2}{c}{\textbf{}}                &      \multicolumn{2}{|c|}{\textbf{Initial}}   &      \multicolumn{2}{|c|}{\textbf{Final}}  &      \multicolumn{2}{|c|}{\textbf{\% Reduction}} \\ \hline
\textbf{Ref} & \textbf{Fig} & \textbf{ Gates} & \textbf{ Levels} & \textbf{ Gates} & \textbf{ Levels} & \textbf{ Gates} & \textbf{ Levels} \\ \hline
\cite{majumder2018experimental} & 26  & 18 & 8  & 4  & 3  & 78 & 63 \\ \hline
\cite{majumder2018experimental} & 27  & 19 & 9  & 5  & 4  & 74 & 56 \\ \hline
\cite{majumder2018experimental} & 29  & 19 & 8  & 5  & 4  & 74 & 50 \\ \hline
\cite{majumder2018experimental} & 25  & 17 & 8  & 5  & 4  & 71 & 50 \\ \hline
\cite{alsina2016experimental}  & 1a  & 12 & 7  & 4  & 2  & 67 & 71 \\ \hline
\cite{majumder2018experimental} & 28  & 20 & 8  & 8  & 4  & 60 & 50 \\ \hline
\cite{sisodia2017design} & 3   & 28 & 22 & 13 & 11 & 54 & 50 \\ \hline
\cite{deffner2017demonstration}  & 5   & 4  & 3  & 2  & 2  & 50 & 33 \\ \hline
\cite{martin2018five}  & 13a & 8  & 6  & 4  & 3  & 50 & 50 \\ \hline
\cite{bikash2018purification}  & 3   & 37 & 21 & 19 & 15 & 49 & 29 \\ \hline
\cite{alsina2016experimental}  & 1b  & 14 & 7  & 8  & 5  & 43 & 29 \\ \hline
\cite{alsina2016experimental}  & 2b  & 20 & 7  & 12 & 5  & 40 & 29 \\ \hline
\cite{deffner2017demonstration}  & 3   & 11 & 5  & 7  & 5  & 36 & 0  \\ \hline
\cite{sisodia2017experimental}  & 4b  & 11 & 7  & 7  & 6  & 36 & 14 \\ \hline
\cite{deffner2017demonstration}  & 6   & 6  & 4  & 4  & 4  & 33 & 0  \\ \hline
\cite{srinivasan2017solving}  & 2   & 13 & 9  & 9  & 7  & 31 & 22 \\ \hline
\cite{satyajit2018nondestructive} & 6   & 23 & 15 & 17 & 11 & 26 & 27 \\ \hline
\cite{alsina2016experimental}  & 2a  & 17 & 7  & 13 & 6  & 24 & 14 \\ \hline
\cite{martin2018five}  & 13b & 10 & 6  & 8  & 5  & 20 & 17 \\ \hline
\cite{sisodia2017experimental}  & 3b  & 11 & 8  & 9  & 8  & 18 & 0  \\ \hline
\cite{li2017approximate}  & 6   & 41 & 27 & 34 & 23 & 17 & 15 \\ \hline
\cite{kalra2017demonstration}  & 2   & 13 & 10 & 11 & 7  & 15 & 30 \\ \hline
\cite{kalra2017demonstration}  & 3   & 17 & 14 & 15 & 11 & 12 & 21 \\ \hline
\cite{ghosh2018automated}  & 4   & 37 & 27 & 33 & 25 & 11 & 7  \\ \hline
\cite{ghosh2018automated}  & 8   & 57 & 38 & 51 & 37 & 11 & 3  \\ \hline
\cite{ghosh2018automated}  & 9   & 19 & 12 & 17 & 12 & 11 & 0  \\ \hline
\cite{vishnu2018experimental} & 6   & 24 & 13 & 22 & 13 & 8  & 0  \\ \hline
\cite{ghosh2018automated}  & 3   & 28 & 17 & 26 & 17 & 7  & 0  \\ \hline
\cite{sisodia2017experimental}  & 8   & 28 & 19 & 26 & 18 & 7  & 5  \\ \hline
\cite{bikash2018purification}  & 6   & 29 & 20 & 27 & 19 & 7  & 5  \\ \hline
\cite{ghosh2018automated}  & 10  & 30 & 19 & 28 & 19 & 7  & 0  \\ \hline
\cite{kamal2017fixed-point} & 1   & 53 & 31 & 50 & 31 & 6  & 0  \\ \hline
\end{tabular}
\caption{Results of the optimization of the quantum circuits which were implemented earlier using IBM quantum processor. Here, the optimization is done using our  algorithm  and by considering the architecture used in QX2}\label{QX2}
\end{center} \end{table}

\begin{table}
{ \footnotesize
\begin{center}
\begin{tabular}{lcc|c|c|c|c|c|c|}
\cline{4-9}
                                            &                                             &                & \multicolumn{2}{c|}{\textbf{Qiskit}} & \multicolumn{2}{c|}{\textbf{RevKit}} & \multicolumn{2}{c|}{\textbf{Improvements (\%)}} \\ \hline
\multicolumn{1}{|l|}{\textbf{Benchmarks}}   & \multicolumn{1}{c|}{\textbf{Initial}} & \textbf{lines} & \textbf{gates}   & \textbf{levels}   & \textbf{gates}   & \textbf{levels}   & \textbf{gates}         & \textbf{levels}        \\ \hline
\multicolumn{1}{|l|}{alu-v1\_28}            & \multicolumn{1}{c|}{37}                     & 5          & 179              & 82          & 51               & 29              & 72                     & 65                     \\ \hline
\multicolumn{1}{|l|}{alu-v4\_37}            & \multicolumn{1}{c|}{37}                     & 4          & 163              & 91          & 48               & 27              & 71                     & 70                     \\ \hline
\multicolumn{1}{|l|}{alu-v0\_27}            & \multicolumn{1}{c|}{36}                     & 3          & 151              & 80          & 47               & 26              & 69                     & 68                     \\ \hline
\multicolumn{1}{|l|}{rd32-v1\_68}           & \multicolumn{1}{c|}{36}                     & 3          & 123              & 70          & 41               & 27              & 67                     & 61                     \\ \hline
\multicolumn{1}{|l|}{alu-v2\_33}            & \multicolumn{1}{c|}{37}                     & 3          & 144              & 79          & 50               & 27              & 65                     & 66                     \\ \hline
\multicolumn{1}{|l|}{alu-v1\_29}            & \multicolumn{1}{c|}{37}                     & 3          & 138              & 73          & 48               & 26              & 65                     & 64                     \\ \hline
\multicolumn{1}{|l|}{a3x\_c}                & \multicolumn{1}{c|}{48}                     & 3          & 143              & 71          & 52               & 39              & 64                     & 45                     \\ \hline
\multicolumn{1}{|l|}{rd32-v0\_66}           & \multicolumn{1}{c|}{34}                     & 3          & 111              & 65          & 41               & 27              & 63                     & 58                     \\ \hline
\multicolumn{1}{|l|}{decod24-v0\_38}        & \multicolumn{1}{c|}{51}                     & 5          & 180              & 99          & 67               & 42              & 63                     & 58                     \\ \hline
\multicolumn{1}{|l|}{mod5mils\_65}          & \multicolumn{1}{c|}{35}                     & 5          & 140              & 79          & 54               & 35              & 61                     & 56                     \\ \hline
\multicolumn{1}{|l|}{one-two-three-v2\_100} & \multicolumn{1}{c|}{69}                     & 5          & 238              & 128         & 95               & 48              & 60                     & 63                     \\ \hline
\multicolumn{1}{|l|}{4gt11\_82}             & \multicolumn{1}{c|}{27}                     & 5          & 136              & 69          & 55               & 33              & 60                     & 52                     \\ \hline
\multicolumn{1}{|l|}{alu-v0\_26}            & \multicolumn{1}{c|}{84}                     & 5          & 295              & 157         & 120              & 72              & 59                     & 54                     \\ \hline
\multicolumn{1}{|l|}{a3x\_d}                & \multicolumn{1}{c|}{44}                     & 5          & 116              & 69          & 48               & 33              & 59                     & 52                     \\ \hline
\multicolumn{1}{|l|}{one-two-three-v3\_101} & \multicolumn{1}{c|}{70}                     & 5          & 291              & 166         & 121              & 74              & 58                     & 55                     \\ \hline
\multicolumn{1}{|l|}{alu-v3\_35}            & \multicolumn{1}{c|}{37}                     & 5          & 113              & 61          & 48               & 27              & 58                     & 56                     \\ \hline
\multicolumn{1}{|l|}{Full\_Adder\_c}        & \multicolumn{1}{c|}{20}                     & 5          & 58               & 38          & 25               & 22              & 57                     & 42                     \\ \hline
\multicolumn{1}{|l|}{mod5d1\_63}            & \multicolumn{1}{c|}{22}                     & 5          & 90               & 49          & 39               & 27              & 57                     & 45                     \\ \hline
\multicolumn{1}{|l|}{mod5d2\_64}            & \multicolumn{1}{c|}{53}                     & 5          & 206              & 119         & 94               & 61              & 54                     & 49                     \\ \hline
\multicolumn{1}{|l|}{4mod5-v1\_22}          & \multicolumn{1}{c|}{21}                     & 5          & 74               & 40          & 35               & 23              & 53                     & 43                     \\ \hline
\multicolumn{1}{|l|}{decod24-v2\_43}        & \multicolumn{1}{c|}{52}                     & 5          & 152              & 80          & 74               & 47              & 51                     & 41                     \\ \hline
\multicolumn{1}{|l|}{alu-v3\_34}            & \multicolumn{1}{c|}{52}                     & 5          & 153              & 88          & 75               & 45              & 51                     & 49                     \\ \hline
\multicolumn{1}{|l|}{X1}                    & \multicolumn{1}{c|}{51}                     & 5          & 127              & 64          & 63               & 30              & 50                     & 53                     \\ \hline
\multicolumn{1}{|l|}{decod24-v1\_41}        & \multicolumn{1}{c|}{85}                     & 5          & 281              & 161         & 149              & 83              & 47                     & 48                     \\ \hline
\multicolumn{1}{|l|}{4mod5-v0\_19}          & \multicolumn{1}{c|}{35}                     & 5          & 98               & 61          & 53               & 32              & 46                     & 48                     \\ \hline
\multicolumn{1}{|l|}{4mod5-v1\_24}          & \multicolumn{1}{c|}{36}                     & 5          & 105              & 55          & 59               & 38              & 44                     & 31                     \\ \hline
\multicolumn{1}{|l|}{4mod5-v1\_23}          & \multicolumn{1}{c|}{69}                     & 5          & 228              & 121         & 130              & 78              & 43                     & 36                     \\ \hline
\multicolumn{1}{|l|}{Full\_Adder\_e}        & \multicolumn{1}{c|}{21}                     & 5          & 58               & 30          & 34               & 18              & 41                     & 40                     \\ \hline
\multicolumn{1}{|l|}{a2x\_c}                & \multicolumn{1}{c|}{31}                     & 5          & 67               & 40          & 40               & 28              & 40                     & 30                     \\ \hline
\multicolumn{1}{|l|}{decod24-v3\_45}        & \multicolumn{1}{c|}{150}                    & 4          & 462              & 275         & 281              & 179             & 39                     & 35                     \\ \hline
\multicolumn{1}{|l|}{17}                    & \multicolumn{1}{c|}{43}                     & 4          & 149              & 91          & 91               & 55              & 39                     & 40                     \\ \hline
\multicolumn{1}{|l|}{4gt5\_76}              & \multicolumn{1}{c|}{91}                     & 5          & 294              & 146         & 182              & 111             & 38                     & 24                     \\ \hline
\multicolumn{1}{|l|}{a2x\_e}                & \multicolumn{1}{c|}{30}                     & 5          & 66               & 38          & 41               & 27              & 38                     & 29                     \\ \hline
\multicolumn{1}{|l|}{alu-v4\_36}            & \multicolumn{1}{c|}{115}                    & 5          & 339              & 185         & 211              & 123             & 38                     & 34                     \\ \hline
\multicolumn{1}{|l|}{hwb4\_49}              & \multicolumn{1}{c|}{233}                    & 5          & 788              & 430         & 494              & 315             & 37                     & 27                     \\ \hline
\multicolumn{1}{|l|}{4gt13\_90}             & \multicolumn{1}{c|}{107}                    & 5          & 309              & 163         & 200              & 116             & 35                     & 29                     \\ \hline
\multicolumn{1}{|l|}{aj-e11\_165}           & \multicolumn{1}{c|}{151}                    & 5          & 448              & 242         & 294              & 182             & 34                     & 25                     \\ \hline
\multicolumn{1}{|l|}{4\_49\_16}             & \multicolumn{1}{c|}{217}                    & 5          & 633              & 332         & 418              & 262             & 34                     & 21                     \\ \hline
\multicolumn{1}{|l|}{7}                     & \multicolumn{1}{c|}{60}                     & 5          & 165              & 97          & 111              & 65              & 33                     & 33                     \\ \hline
\multicolumn{1}{|l|}{4mod5-v0\_20}          & \multicolumn{1}{c|}{20}                     & 5          & 46               & 26          & 31               & 20              & 33                     & 23                     \\ \hline
\multicolumn{1}{|l|}{rd32\_270}             & \multicolumn{1}{c|}{84}                     & 5          & 216              & 114         & 155              & 98              & 28                     & 14                     \\ \hline
\multicolumn{1}{|l|}{alu-v2\_32}            & \multicolumn{1}{c|}{163}                    & 5          & 423              & 221         & 309              & 175             & 27                     & 21                     \\ \hline
\multicolumn{1}{|l|}{4gt11\_84}             & \multicolumn{1}{c|}{18}                     & 5          & 34               & 20          & 25               & 12              & 26                     & 40                     \\ \hline
\multicolumn{1}{|l|}{4mod7-v0\_94}          & \multicolumn{1}{c|}{162}                    & 5          & 399              & 230         & 299              & 184             & 25                     & 20                     \\ \hline
\multicolumn{1}{|l|}{Full\_Adder\_d}        & \multicolumn{1}{c|}{22}                     & 5          & 49               & 32          & 37               & 24              & 24                     & 25                     \\ \hline
\multicolumn{1}{|l|}{4mod5-v0\_18}          & \multicolumn{1}{c|}{69}                     & 4          & 173              & 102         & 139              & 91              & 20                     & 11                     \\ \hline
\multicolumn{1}{|l|}{alu-v2\_31}            & \multicolumn{1}{c|}{451}                    & 5          & 1163             & 625         & 942              & 578             & 19                     & 8                      \\ \hline
\multicolumn{1}{|l|}{mini-alu\_167}         & \multicolumn{1}{c|}{288}                    & 4          & 712              & 369         & 577              & 353             & 19                     & 4                      \\ \hline
\multicolumn{1}{|l|}{4gt10-v1\_81}          & \multicolumn{1}{c|}{148}                    & 5          & 348              & 183         & 286              & 174             & 18                     & 5                      \\ \hline
\multicolumn{1}{|l|}{Toffoli\_e}            & \multicolumn{1}{c|}{17}                     & 3          & 23               & 14          & 19               & 12              & 17                     & 14                     \\ \hline
\multicolumn{1}{|l|}{4gt13-v1\_93}          & \multicolumn{1}{c|}{68}                     & 4          & 104              & 51          & 86               & 48              & 17                     & 6                      \\ \hline
\multicolumn{1}{|l|}{4gt13\_91}             & \multicolumn{1}{c|}{103}                    & 4          & 227              & 131         & 199              & 111             & 12                     & 15                     \\ \hline
\multicolumn{1}{|l|}{ex-1\_166}             & \multicolumn{1}{c|}{19}                     & 4          & 26               & 18          & 23               & 15              & 12                     & 17                     \\ \hline
\end{tabular}
\end{center} }
\caption{Results of the optimization of the reversible circuits.  The optimization is done using our  algorithm  and by considering the architecture used in QX4. Here, initial circuits correspond to the Clifford+T circuits for various functions available in \cite{benchmarks}. No restriction on application of CNOT gates is applied in the initial circuits. Considering the restrictions on CNOT gates implied by the architecture of QX4, the columns under Qiskit \cite{qiskit} is obtained. Finally, those same circuits are optimized using our algorithm and the corresponding results are shown in the columns under RevKit \cite{RevKit}, as the RevKit platform is used by us.} \label{QX4b}

\end{table}


\section{Conclusion \label{sec:Conclusion}}

In this paper, we have shown that quantum circuits which have been
designed until now, for the implementations in IBM quantum computers
can be optimized in terms of gate count and the number of levels. It's
further established that the reduction in these measures of circuit
costs leads to improvement in the fidelity of the output state, and
thus reduces the effect of noise or equivalently increases the accuracy of the quantum computation. As the effect of noise generally
leads to an evolution of a quantum state towards a classical state,
the reduction of noise through optimization of quantum circuits are
expected to lead to clearer manifestation of quantum features. This
has been clearly seen in the context of 3-qubit Mermin inequality, where the amount of violation of Mermin inequality can be quantified by 
$\langle M_3 \rangle-2$, a greater value of which would correspond to the stronger signature of departure from the classical world. The optimized circuit has yielded a value of this quantity as 1.116 in contrast to the earlier reported value of 0.85, clearly indicating a stronger signature of non-locality. Thus, the optimized circuit illustrate the quantum feature
in a more profound manner. Also, as expected the fidelity of the optimized 
circuits is found to be higher than their non-optimized counter
parts. This establishes that to obtain best results using IBM quantum
computers, one has to use our approach and keeping this fact in mind,
we conclude this paper by noting that this work is expected to influence
a large number of future works involving IBM quantum experience by
providing a tool for obtaining best results. Of course, the results reported here are restricted to 5-qubit quantum computers, but the method used is general and the program can be scaled up for 16 qubit, 20 qubit and other bigger computers, too. The results
for such systems will be reported elsewhere. 

\section*{Acknowledgments}

A.S. gratefully acknowledge University of Science and Technology of China (USTC), Hefei, P. R. C. for providing support. A.A.A. is grateful for the support of Coordena\c{c}\~ao de Aperfei\c{c}oamento de Pessoal de N\'{\i}vel Superior - Brasil (CAPES) - Finance Code 001, and he also thanks Universidade Estadual Paulista (UNESP), Faculdade de Engenharia de Ilha Solteira, C\^ampus de Ilha Solteira.  G.W.D. and A.A.A. also acknowledge the support of the Natural Sciences and Engineering Research Council of Canada (NSERC). A.P. thanks the Department of Science and Technology (DST), India, for support provided through the DST project No. EMR/2015/000393. 

\bibliographystyle{Final}
\bibliography{mitali1}

\begin{thebibliography}{10}
\expandafter\ifx\csname urlstyle\endcsname\relax
  \providecommand{\doi}[1]{doi:\discretionary{}{}{}#1}\else
  \providecommand{\doi}{doi:\discretionary{}{}{}\begingroup
  \urlstyle{rm}\Url}\fi
\providecommand{\bibAnnoteFile}[1]{%
  \IfFileExists{#1}{\begin{quotation}\noindent\textsc{Key:} #1\\
  \textsc{Annotation:}\ \input{#1}\end{quotation}}{}}
\providecommand{\bibAnnote}[2]{%
  \begin{quotation}\noindent\textsc{Key:} #1\\
  \textsc{Annotation:}\ #2\end{quotation}}

\bibitem{grover1997quantum}
Grover, L.~K.: Quantum mechanics helps in searching for a needle in a haystack.
  Physical Review Letters \textbf{79}, 325 (1997)
\bibAnnoteFile{grover1997quantum}

\bibitem{shor1999polynomial}
Shor, P.~W.: Polynomial-time algorithms for prime factorization and discrete
  logarithms on a quantum computer. SIAM review \textbf{41}, 303--332 (1999)
\bibAnnoteFile{shor1999polynomial}

\bibitem{bennett1993teleporting}
Bennett, C.~H., Brassard, G., Cr{\'e}peau, C., et~al.: Teleporting an unknown
  quantum state via dual classical and {Einstein-Podolsky-Rosen} channels.
  Physical Review Letters \textbf{70}, 1895 (1993)
\bibAnnoteFile{bennett1993teleporting}

\bibitem{shenoy2017quantum}
Shenoy-Hejamadi, A., Pathak, A., Radhakrishna, S.: Quantum cryptography: Key
  distribution and beyond. Quanta \textbf{6} (2017)
\bibAnnoteFile{shenoy2017quantum}

\bibitem{IBMQ}
{IBM Q}. \url{https://www.research.ibm.com/ibm-q/}. Accessed: 2018-11-05
\bibAnnoteFile{IBMQ}

\bibitem{sisodia2017experimental}
Sisodia, M., Shukla, A., Pathak, A.: Experimental realization of nondestructive
  discrimination of {Bell} states using a five-qubit quantum computer. Physics
  Letters A \textbf{381}, 3860--3874 (2017)
\bibAnnoteFile{sisodia2017experimental}

\bibitem{sisodia2017design}
Sisodia, M., Shukla, A., Thapliyal, K., Pathak, A.: Design and experimental
  realization of an optimal scheme for teleportation of an n-qubit quantum
  state. Quantum Information Processing \textbf{16}, 292 (2017)
\bibAnnoteFile{sisodia2017design}

\bibitem{YZ2017optimization}
Yal{\c{c}}{\i}nkaya, {\.I}., Gedik, Z.: Optimization and experimental
  realization of the quantum permutation algorithm. Physical Review A
  \textbf{96}, 062339 (2017)
\bibAnnoteFile{YZ2017optimization}

\bibitem{BBP2017experimental}
Behera, B.~K., Banerjee, A., Panigrahi, P.~K.: Experimental realization of
  quantum cheque using a five-qubit quantum computer. Quantum Information
  Processing \textbf{16}, 312 (2017)
\bibAnnoteFile{BBP2017experimental}

\bibitem{AL2016experimental}
Alsina, D., Latorre, J.~I.: Experimental test of {Mermin} inequalities on a
  five-qubit quantum computer. Physical Review A \textbf{94}, 012314 (2016)
\bibAnnoteFile{AL2016experimental}

\bibitem{shukla2018complete}
Shukla, A., Sisodia, M., Pathak, A.: Complete characterization of the
  single-qubit quantum gates used in the {IBM} quantum processors. arXiv
  preprint arXiv:1805.07185  (2018)
\bibAnnoteFile{shukla2018complete}

\bibitem{dueck_DSD_2018}
Dueck, G.~W., Pathak, A., Rahman, M.~M., Shukla, A., Banerjee, A.: Optimization
  of circuits for {IBM}'s five-qubit quantum computers. in: 2018 21st Euromicro
  Conference on Digital System Design (DSD) pp. 680--684 IEEE (2018)
\bibAnnoteFile{dueck_DSD_2018}

\bibitem{IBMQ2_info}
{IBM Q5 Yorktown}.
  \url{https://github.com/Qiskit/ibmq-device-information/tree/master/backends/yorktown/V1}.
  Accessed: 2018-11-05
\bibAnnoteFile{IBMQ2_info}

\bibitem{IBMQ4_info}
{IBM Q5 Tenerife}.
  \url{https://github.com/Qiskit/ibmq-device-information/tree/master/backends/tenerife/V1}.
  Accessed: 2018-11-05
\bibAnnoteFile{IBMQ4_info}

\bibitem{lanyon2014experimental}
Lanyon, B., Zwerger, M., Jurcevic, P., et~al.: Experimental violation of
  multipartite {Bell} inequalities with trapped ions. Physical Review Letters
  \textbf{112}, 100403 (2014)
\bibAnnoteFile{lanyon2014experimental}

\bibitem{zhao2003experimental}
Zhao, Z., Yang, T., Chen, Y.-A., et~al.: Experimental violation of local
  realism by four-photon {Greenberger-Horne-Zeilinger} entanglement. Physical
  Review Letters \textbf{91}, 180401 (2003)
\bibAnnoteFile{zhao2003experimental}

\bibitem{benchmarks}
{Reversible and Quantum Benchmark Circuits}.
  \url{https://github.com/QBenchmark/benchmarks}
\bibAnnoteFile{benchmarks}

\bibitem{majumder2018experimental}
Majumder, A., Kumar, A.: Experimental demonstration of non-destructive
  discrimination of arbitrary set of orthogonal quantum states using 5-qubit
  ibm quantum computer on cloud. arXiv preprint arXiv:1803.06311  (2018)
\bibAnnoteFile{majumder2018experimental}

\bibitem{RevLib}
Gro{\ss}e, D., Wille, R., Dueck, G.~W., Drechsler, R.: Exact synthesis of
  elementary quantum gate circuits for reversible functions with don't cares.
  in: {Int'l Symp. on {M}ulti-{V}alued {L}ogic} pp. 214--219 (2008)
\bibAnnoteFile{RevLib}

\bibitem{alsina2016experimental}
Alsina, D., Latorre, J.~I.: Experimental test of {Mermin} inequalities on a
  five-qubit quantum computer. Physical Review A \textbf{94}, 012314 (2016)
\bibAnnoteFile{alsina2016experimental}

\bibitem{deffner2017demonstration}
Deffner, S.: Demonstration of entanglement assisted invariance on ibm's quantum
  experience. Heliyon \textbf{3}, e00444 (2017)
\bibAnnoteFile{deffner2017demonstration}

\bibitem{martin2018five}
Garc\'ia-Mart\'in, D., Sierra, G.: Five experimental tests on the 5-qubit {IBM}
  quantum computer. Journal of Applied Mathematics and Physics  (2018)
\bibAnnoteFile{martin2018five}

\bibitem{bikash2018purification}
Behera, B.~K., Seth, S., Das, A., Panigrahi, P.~K.: Demonstration of
  entanglement purification and swapping protocol to design quantum repeater in
  {IBM} quantum computer. arXiv preprint arXiv:1712.00854v2  (2018)
\bibAnnoteFile{bikash2018purification}

\bibitem{srinivasan2017solving}
Srinivasan, K., Behera, B.~K., Panigrahi, P.~K.: Solving linear systems of
  equations by gaussian elimination method using grover's search algorithm: An
  {IBM} quantum experience. arXiv preprint arXiv:1801.00778  (2017)
\bibAnnoteFile{srinivasan2017solving}

\bibitem{li2017approximate}
Li, R., Alvarez-Rodriguez, U., Lamata, L., Solano, E.: Approximate quantum
  adders with genetic algorithms: an {IBM} quantum experience. Quantum
  Measurements and Quantum Metrology \textbf{4}, 1--7 (2017)
\bibAnnoteFile{li2017approximate}

\bibitem{vishnu2018experimental}
Vishnu, P., Joy, D., Behera, B.~K., Panigrahi, P.~K.: Experimental
  demonstration of non-local controlled-unitary quantum gates using a
  five-qubit quantum computer. Quantum Information Processing \textbf{17}, 274
  (2018)
\bibAnnoteFile{vishnu2018experimental}

\bibitem{kalra2017demonstration}
Kalra, A.~R., Prakash, S., Behera, B.~K., Panigrahi, P.~K.: Demonstration of
  the quantum no-hiding theorem in a category theoretic framework: An {IBM}
  quantum experience  (2017)
\bibAnnoteFile{kalra2017demonstration}

\bibitem{satyajit2018nondestructive}
Satyajit, S., Srinivasan, K., Behera, B.~K., Panigrahi, P.~K.: Nondestructive
  discrimination of a new family of highly entangled states in {IBM} quantum
  computer. Quantum Information Processing \textbf{17}, 212 (2018)
\bibAnnoteFile{satyajit2018nondestructive}

\bibitem{kamal2017fixed-point}
Gurnani, K., Behera, B.~K., Panigrahi, P.~K.: Demonstration of optimal
  fixed-point quantum search algorithm in {IBM} quantum computer. arXiv
  preprint arXiv:1712.10231v1  (2017)
\bibAnnoteFile{kamal2017fixed-point}

\bibitem{ghosh2018automated}
Ghosh, D., Agarwal, P., Pandey, P., Behera, B.~K., Panigrahi, P.~K.: Automated
  error correction in {IBM} quantum computer and explicit generalization.
  Quantum Information Processing \textbf{17}, 153 (2018)
\bibAnnoteFile{ghosh2018automated}

\bibitem{qiskit}
{Qiskit - An open-source framework for working with noisy intermediate-scale
  quantum computers (NISQ) at the level of pulses, circuits, and algorithms.}
  \url{https://github.com/Qiskit}
\bibAnnoteFile{qiskit}

\bibitem{RevKit}
Soeken, M., Frehse, S., Wille, R., Drechsler, R.: {RevKit:} a toolkit for
  reversible circuit design. in: Proceedings of the International Symposium on
  Multiple-Valued Logic (2008). {RevKit} is available at
  https://github.com/msoeken/cirkit
\bibAnnoteFile{RevKit}

\end{thebibliography}
\end{document}